\documentclass[journal]{IEEEtran}
\usepackage{url}

\begin{document}

\title{A 3D Game Theoretical Framework for the Evaluation of Unmanned Aircraft Systems Airspace Integration Concepts}

\author{Negin~Musavi,~\IEEEmembership{}
        Ayman~Manzoor,~\IEEEmembership{}
        and~Yildiray~Yildiz,~\IEEEmembership{Senior~Member,~IEEE}

\thanks{N. Musavi, A. Manzoor, and Y. Yildiz are with the Department
of Mechanical Engineering, Bilkent University, Ankara, 06800 Turkey (e-mail: \protect\url{nmusavi2@illinois.edu}; \protect\url{ayman.manzoor@ug.bilkent.edu.tr}; \protect\url{yyildiz@bilkent.edu.tr}).}
}

\maketitle

\begin{abstract}
Predicting the outcomes of integrating Unmanned Aerial Systems (UAS) into the National Airspace System (NAS) is a complex problem which is required to be addressed by simulation studies before allowing the routine access of UAS into the NAS. This paper focuses on providing a 3-dimensional (3D) simulation framework using a game theoretical methodology to evaluate integration concepts using scenarios where manned and unmanned air vehicles co-exist. In the proposed method, human pilot interactive decision making process is incorporated into airspace models which can fill the gap in the literature where the pilot behavior is generally assumed to be known a priori. The proposed human pilot behavior is modeled using dynamic level-k reasoning concept and approximate reinforcement learning. The level-k reasoning concept is a notion in game theory and is based on the assumption that humans have various levels of decision making. In the conventional ``static" approach, each agent makes assumptions about his or her opponents and chooses his or her actions accordingly. On the other hand, in the dynamic level-k reasoning, agents can update their beliefs about their opponents and revise their level-k rule. In this study, Neural Fitted Q Iteration, which is an approximate reinforcement learning method, is used to model time-extended decisions of pilots with 3D maneuvers. An analysis of UAS integration is conducted using an example 3D scenario in the presence of manned aircraft and fully autonomous UAS equipped with sense and avoid algorithms.
\end{abstract}



\section{Introduction}\label{Introduction}

\IEEEPARstart{A}{lthough} unmanned aircraft systems (UAS) have operational and cost advantages over manned aircraft in many applications, they do not have routine access to the National Airspace (NAS). The aviation industry, being very sensitive to safety, needs strong evidence that the UAS integration will not have any negative impact on the existing airspace system in terms of safety \cite{US_UAV:13, European_UAV:13} before they are granted routine access into the NAS. Until technologies, standards, and procedures for a safe integration of UAS into the airspace are matured, there will not be enough data accumulated about the issue and it will be hard to predict the effectiveness of the related technologies and concepts. Although research efforts exist to develop a safe and efficient real test environment for UAS integration \cite{NASA:12}, flight tests are expensive and experimental failures can cause severe economic loss. Therefore, employing simulations is currently the most efficient way to understand the effects of UAS integration on the air traffic system \cite{MITRE:14}. These simulation studies need to be conducted with hybrid airspace system (HAS) models, where manned and unmanned vehicles coexist.

HAS models in the literature are generally based on the assumption that the pilots of manned aircraft always behave as expected, without deviating from ideal behavior \cite{Maki:12}-\cite{Billingsley:06}. Most of the existing HAS models are designed to evaluate and test the performance of collision avoidance systems in single encounter scenarios in which the intruder (generally a manned aircraft) has a pre-defined behavior with no consideration of the decision making process of the pilot. These models are valuable and essential at the initial stages of evaluating a new method but it is not realistic to expect that the pilot, as a decision maker, will always behave deterministically and in a pre-defined manner. It is not always predictable, for example, how pilots will respond to the traffic control alert system (TCAS) \cite{Salas:10}. TCAS is an on-board collision avoidance system which observes and tracks surrounding air traffic, detects conflicts and suggests avoidance maneuvers to the pilots. In recent studies, it was shown that only 13\% of pilot responses match the deterministic pilot model that was assumed for TCAS development \cite{Lee:11}, \cite{Kuchar:07}. Therefore, incorporating human decision-making processes in HAS models has a strong potential to improve the predictive power of these models.

In prior works \cite{Negin:16}, \cite{Musavi:16}, authors have created HAS models with human decision making models, inspired by a game theoretical methodology known as semi-network-form games \cite{Lee:11}, where the pilot behavior was not assumed to be known a priori but obtained using 1) the level-k reasoning concept which is a game theoretical approach used to model multiple strategic player interactions, where it is assumed that humans have various levels of reasoning, level-0 being the lowest level, and 2) reinforcement learning, which helps model time extended decisions as opposed to assuming one-shot decision making. Although these studies introduced one of the very first examples of HAS models where several decision makers can be modeled simultaneously in a time-extended manner, they had two limitations: First, HAS models were developed for a 2-dimensional (2D) airspace. Second, the policies, i.e. maps from observation spaces to action spaces, obtained for the decision makers remain unchanged during their interaction. In the proposed framework, these limitations are removed and a 3D HAS model is introduced where the strategic decision makers can modify their policies during interactions between each other. Therefore, compared to \cite{Negin:16}, \cite{Musavi:16} a much larger class of interactions can be modeled. 

It is shown in the literature that 1) in repeated strategic interactions, where agents consider other agents' possible actions before determining their own, agents with different cognitive abilities change their behavior during the interaction \cite{Gill:14} and 2) there is a positive relationship between cognitive ability and reasoning levels \cite{Gill:12}, \cite{Gill:14}. These observations lead to agents with different levels of reasoning who can observe their opponents' behavior during repeated interactions, update their beliefs on their opponents' reasoning level and change their own level-k rule against them. In \cite{Gill:12} and \cite{Gill:14}, a systematic level-k structure is introduced where players can update their beliefs about their opponents, and switch their own level rule up one level during their interactions. There are also other level-k rule learning models in the literature such as the ones presented in \cite{Chong:16} and \cite{Ho:13}, where the agent levels can reach up to infinity. This is not a problem for the applications investigated in \cite{Chong:16} and \cite{Ho:13}, in which obtaining level-k rules (k=0,1,2,...,$\infty$) are straight forward and has an analytical solution. Since it is computationally expensive to obtain higher levels, and in certain experimental studies it is shown that humans in general have a maximum reasoning level of 2 \cite{Costa:95}, the existing level-k rule learning methods may not be suitable for the application considered in this work where more than 188 decision makers are modeled simultaneously in a time extended manner. Here, we propose a simpler method for modeling level-k rule updates during interactions by a) limiting the levels up to 2 and b) allowing rule updates only if a trajectory conflict is observed. 

Different from the 2D HAS model developed in \cite{Negin:16}, \cite{Musavi:16}, in this study, the game theoretical modeling framework is developed for a 3D HAS model which allows to cover a much larger class of integration scenarios. The reinforcement learning algorithm used in the authors' earlier works \cite{Li:01}, \cite{Yildiz:01}, \cite{Backhaus:01} employ tables to store the Q values of all state (location of the intruder, approach angle of the intruder, best trajectory action, best destination action and previous action)-action (turn left, turn right, go straight) pairs, which define how preferable it is to take a certain action given the observations/states. This poses a challenge for the application of the method to systems with a large numbers of state-action pairs such as the proposed 3D HAS model in this study. To circumvent this issue, Neural Fitted Q Iteration (NFQ) method \cite{Rie:05}, \cite{Rie:07} and \cite{Rie:11}, an approximate reinforcement learning algorithm, is utilized. Approximate reinforcement learning methods use function approximators to represent the Q value function \cite{Volo:15}. In other words, instead of saving Q values for each state-action pair, Q value function is approximated by a function approximator. In the case of NFQ, a neural network is used as the function approximator. NFQ approach also allows using a continuous observation space, which also contributes to obtain a more precise definition of the agents' observations, compared to conventional approaches, where a discretized observation space is required.

In the simulations, pilot models that are obtained using the proposed game theoretical modeling framework are used in complex scenarios, where UAS and manned aircraft co-exist, to analyze the probable outcomes of HAS interactions. HAS scenarios contain interacting humans (pilots) who also interact with multiple UAS with their own sense and avoid (SAA) systems. It is noted that automation algorithms other than SAA systems, such as TCAS, and possible air traffic management instructions can also be incorporated into the proposed framework. During the simulations, UAS fly autonomously based on pre-programmed flight plans but they can deviate from their plans to resolve a possible conflict with the help of their SAA algorithms. In these simulations, as an example to demonstrate how the proposed framework can be utilized, the effect of responsibility assignment for conflict resolutions on the safety and performance of the HAS is analyzed (see \cite{NASA:12} for the importance of these variables and responsibility assignment for UAS integration.). 

The organization of the paper is as follows: In section \ref{UAS Integration Scenario}, the HAS scenario for UAS integration into the NAS is described in detail. In section \ref{Pilot Decision Model}, the proposed pilot decision modeling method is explained. In section \ref{Simulation Results and Discussion}, simulation results are provided. Finally, conclusions are given in section \ref{Conclusion}.

\section{UAS Integration Scenario}\label{UAS Integration Scenario}
\begin{figure}[htb]
	\centering	
	\includegraphics[width=8cm]{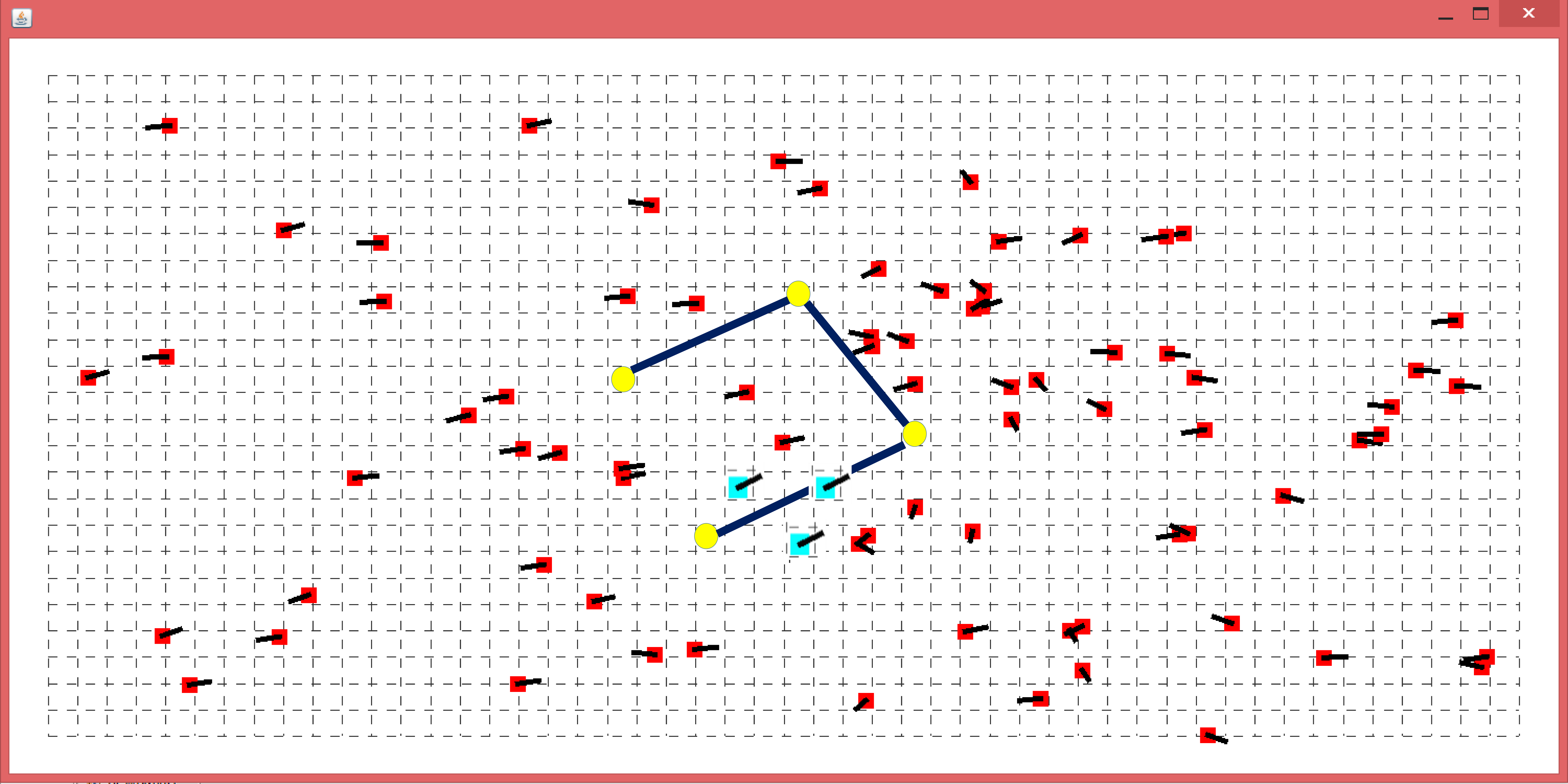}
	\caption{2D snapshot of the airspace scenario in the simulation platform.}
	\label{f:HAS}
\end{figure}

In order to evaluate the possible outcomes of integrating Unmanned Aircraft Systems (UAS) in to the National Air Space (NAS), a Hybrid Air Space (HAS) scenario, where manned and unmanned aircraft co-exist, is designed and explained in this section. The scenario consists of 188 manned aircraft and 3 UAS. The size of the airspace is $600km\times300km (horizontal)\times45000ft (altitude)$. The initial positions, velocities, headings and altitudes of the aircraft are obtained from Flightradar24 website which provides live air traffic data (http://www.flightradar24.com). The data is collected from the air traffic volume on Colorado, USA on March 11, 2015. The manned aircraft in the scenario execute maneuvers based on the pilot model obtained using a combination of reinforcement learning and level-k reasoning, the details of which are explained in Section \ref{Pilot Decision Model}. Multiple UAS are randomly located in the airspace and move based on their pre-programmed flight plan from one waypoint to another. Figure~\ref{f:HAS} shows a snapshot of the scenario with multiple manned aircraft and three UAS moving through their multiple waypoints. All aircraft whether manned or unmanned are flying at different altitudes and this snapshot depicts a 2D projection of their configuration, on the horizontal plane. The red squares correspond to manned aircraft and the cyan squares correspond to UAS, which are flying at different altitudes. All aircraft, manned or unmanned, have continuous dynamics, which are provided in the following sections. Yellow circles show the predetermined waypoints that the UAS with the highest altitude is required to pass. The waypoints of the other two UAS are not shown in this snapshot. The black lines passing through the waypoints show the predetermined path of one of the UAS. It is noted that the UAS do not follow this path exactly since it needs to deviate from its original trajectory to avoid possible conflicts using an on-board Sense and Avoid (SAA) algorithm, which is obtained from \cite{Fasano:08} and \cite{Mujumdar:11}.

In the scenario, it is assumed that each aircraft is able to receive the surrounding traffic information using Automatic Dependent Surveillance-Broadcast (ADS-B) technology. ADS-B technology can provide an own-ship aircraft the identification, position and velocity information of surrounding aircraft that are also equipped with ADS-B.

\subsection{UAS Conflict Detection And Avoidance Logic}\label{UAS Conflict Detection And Avoidance Logic}
UAS fly according to their pre-programmed flight plans, an example of which is marked by yellow shown in Fig.~\ref{f:HAS}. UAS are assumed to have the dynamics of RQ-4 Global Hawk with an operation speed of $340knots$ \cite{Dalamagkidis:09}. UAS are also equipped with SAA systems which enable them to detect trajectory conflicts and to initiate evasive maneuvers, if necessary. If no conflict is detected, UAS continue to follow their mission plan. Either receiving a conflict resolution command from the SAA system or flying based on their pre-defined flight plan, UAS always receives a velocity command during the flight. The UAS velocity vector variation is modeled as first order dynamics with a time constant of $1s$ \cite{Mujumdar:11} which is represented as:
\begin{equation}
\dot{\vec{v}}= -(\vec{v}-\vec{v}_{d}),
\end{equation}
where $\vec{v}$ and $\vec{v}_{d}$ are the current and the desired/commanded velocity vectors, respectively. The two SAA logics that are utilized in this study are developed by Fasano \textit{et al.} \cite{Fasano:08}, which is referred as SAA1, and Mujumdar \textit{et al.} \cite{Mujumdar:11}, which is referred as SAA2. Both of the SAA logics contain two phases; a \textit{conflict detection} phase and a \textit{conflict resolution} phase. The conflict detection phase is the same for both SAA1 and SAA2. A conflict is detected if the minimum distance between the UAS and the intruder aircraft is calculated to be less than a minimum required distance, $R$, during a predefined time interval. The minimum distance is calculated by projecting the trajectories of the UAS and the intruder aircraft in time. Once the conflict is detected, SAA1 and SAA2 suggest their own velocity adjustment commands in order to resolve the conflict. The velocity adjustment command of the SAA1 and SAA2 logics, $\vec{v}_{A}^{d1}$ and  $\vec{v}_{A}^{d2}$, are given in the equations below

\begin{equation}
\label{e:function}
\vec{v}_{A}^{d1} =
\left[ \frac{v_{AB}\cos(\eta-\zeta)}{\sin(\zeta)}[\sin(\eta)\frac{\vec{v}_{AB}}{v_{AB}}-\sin(\eta-\zeta)\frac{\vec{r}}{\|\vec{r}\|}] \right]+\vec{v}_{B}
\end{equation}

\begin{equation}
\label{e:function}
\vec{v}_{A}^{d2} =
\frac{-\vec{v}_{A}(\frac{\vec{r}_{0}.\vec{v}_{AB}}{\|\vec{v}_{AB}\|})-(R-\|\vec{r}_{m}\|)\frac{\vec{r}_{m}}{\|\vec{r}_{m}\|}}{\|-\vec{v}_{A}(\frac{\vec{r}_{0}.\vec{v}_{AB}}{\|\vec{v}_{AB}\|})-(R-\|\vec{r}_{m}\|)\frac{\vec{r}_{m}}{\|\vec{r}_{m}\|}\|}
\end{equation}
\\
where, $\vec{v_{A}}$ and $\vec{v_{B}}$ refer to the velocity vectors of the UAS and the intruder. $\vec{r}$ and $\vec{v_{AB}}$ denote the relative position and velocity between the UAS and the intruder, respectively. $\zeta$ is the angle between $\vec{r}$ and $\vec{v_{AB}}$ and $\eta$ is calculated as $\eta=\sin^{-1}\frac{R}{\|\vec{r}\|}$. $\vec{r}_{0}$ refers to the initial relative position vector between the UAS and the intruder. If multiple conflicts are detected, UAS start an evasive maneuver to resolve the conflict that is predicted to happen earliest. The velocity adjustment suggested by the SAA1 logic guarantees minimum deviation from the trajectory, while in the case of the SAA2 logic, UAS moves to resolve the conflict until it retains the minimum safe distance with the intruder.  

\subsection{Manned Aircraft}\label{Manned Aircraft}
All manned aircraft are assumed to be in their en-route phase of travel with constant speed, $v$, in the range of $[150-550] knots$. Pilots may decide to change the heading angle for $\pm45^{\circ}$, or change the pitch angle for $\pm10^{\circ}$, or may decide to keep both the heading and pitch angles unchanged. Once the pilot gives a heading or pitch command, the aircraft moves to the desired heading and pitch, $\psi_{d}$ and $\theta_{d}$, in the constant speed mode where, the heading and pitch change is modeled by first order dynamics with the standard rate turn: a turn in which an aircraft changes its heading at a rate of $3^{\circ}$ per second ($360^{\circ}$ in 2 minutes) \cite{pilot-handbook:08}. This is modeled approximately by a first order dynamics with a time constant of $10s$ ($45 \times (1-1/e)/3 \approx 10$). Therefore, the aircraft heading and pitch angle dynamics can be given as
\begin{equation}
\dot{\psi}= -\frac{1}{10}\times(\psi-\psi_{d})
\end{equation}
\begin{equation}
\dot{\theta}= -\frac{1}{10}\times(\theta-\theta_{d})
\end{equation}
and the velocity, $\vec{v}=(v_{x},v_{y},v_{z})$, is then obtained as:
\begin{equation}
\\v_{x}= \|\vec{v}\|\sin\psi\cos\theta.
\end{equation}
\begin{equation}
\\v_{y}= \|\vec{v}\|\cos\psi\cos\theta.
\end{equation}
\begin{equation}
\\v_{z}= \|\vec{v}\|\sin\theta.
\end{equation}

\section{Pilot Decision Model}\label{Pilot Decision Model}
The proposed model for pilot decision making in this study is formed by combining two methodologies: dynamic level-k reasoning and neural fitted Q iteration (NFQ), which is an approximate reinforcement learning algorithm. A level-k-type model is trained by assigning level-(k-1)-type behavior to all of the agents (manned aircraft) except the one that is being trained. The trainee learns to react as best as he/she can in this environment using NFQ. Thus, the resulting behavior becomes a level-k type. This process starts with training a level-1 type behavior and continues until the highest desired level is reached. Once all of the desired levels are obtained, the training stage ends and, in the simulation stage, the obtained level-k reaction models are used in the airspace scenario explained in section \ref{UAS Integration Scenario} where both manned aircraft and UAS co-exist. In the simulation, certain proportions of level-0, level-1 and level-2 behavior type are assigned to the manned aircraft. It is noted that each of level-1 and level-2 agents can change their level-k behavior type based on Dynamic level-k reasoning method after observing their intruder's behavior.

\subsection{Dynamic Level-k Reasoning}\label{Dynamic Level-k Reasoning}
Level-k reasoning is a game theoretical model where the main idea is that humans have various levels of reasoning in their decision-making process \cite{Chong:16}. It has been observed that reasoning levels are related to the cognitive abilities of humans \cite{Gill:16}. The level hierarchy is iteratively defined such that the level-k rule is a best response to the level-(k-1) rule.  A level-1 decision maker (DM), for example, assumes that the other agents in the scenario are level-0 and takes actions accordingly to provide the best response. A level-2 DM takes actions to give the best response to other DMs that have level-1 reasoning and so on. From a modeling standpoint, the level-0 rule represents an initial point from which more sophisticated rules can be obtained iteratively. A level-0 rule represents a ``nonstrategic" DM who does not take into account other DMs' possible moves when choosing his/her own actions. This behavior can also be considered as ``reflexive" since it only reacts to the immediate observations. In this study, a level-0 pilot flies an aircraft with constant heading and pitch angles starting from its initial position toward its destination.

\begin{figure}[htb]
	\centering	
	\includegraphics[width=8cm]{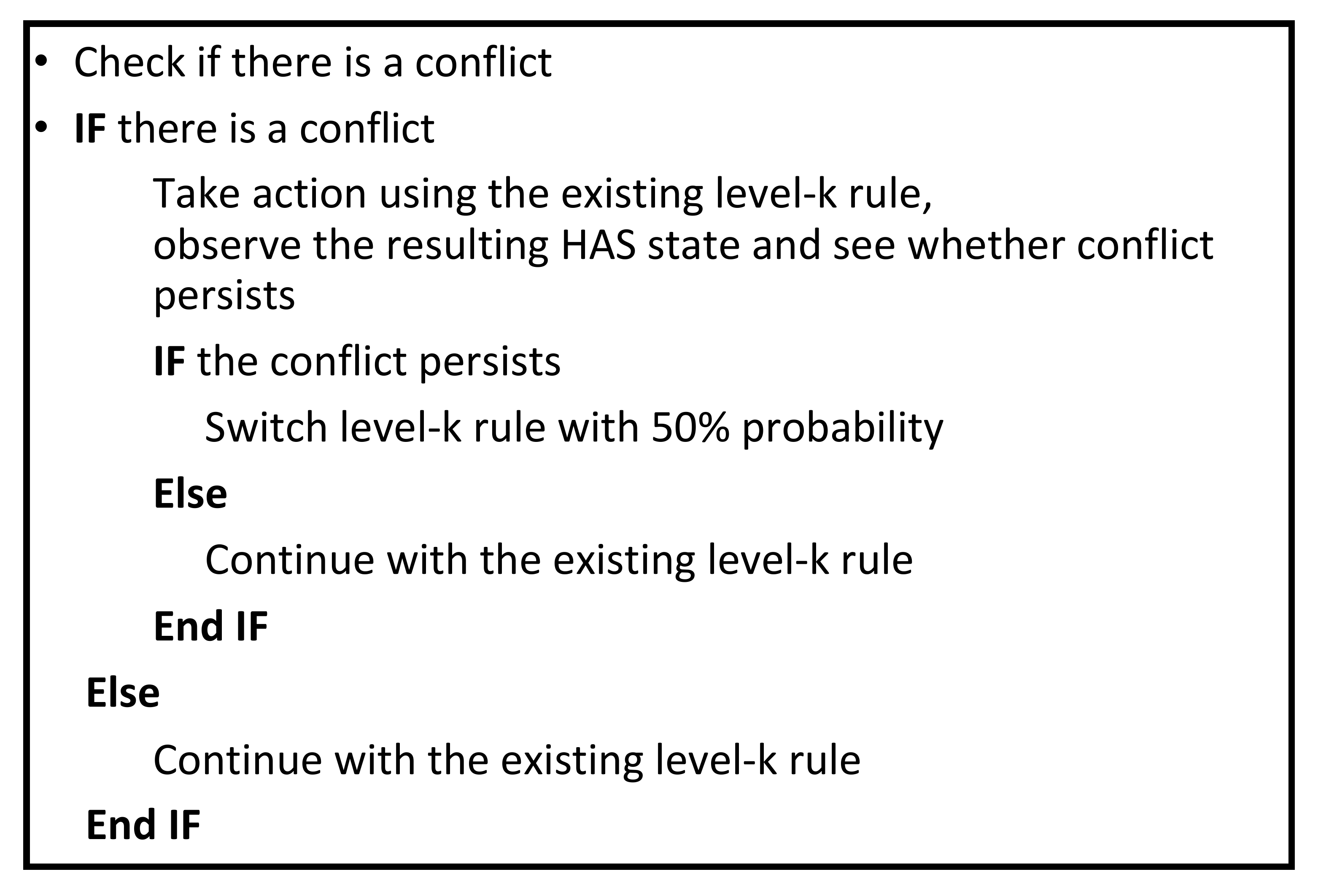}
	\caption{Dynamic level-k reasoning pseudo-code.}
	\label{f:PSC}
\end{figure}

In its conventional form, level-k reasoning help model the interactions between the DMs where a level-k DM assumes that the other DMs have level-(k-1) reasoning. Although this approach proved to be successful in modeling  short term or one-shot interactions, it misses the point that agents, during their interactions, may update their assumptions about the other agents and in turn update their own behavior. To remedy this problem, we introduce a closed loop algorithm which allows the agents to dynamically update their reasoning levels if a trajectory conflict is detected. This algorithm is explained in Fig.~\ref{f:PSC}, where a pseudo-code is provided.

\subsection{Neural Fitted Q Iteration}\label{Neural Fitted Q Iteration}
Reinforcement learning is a mathematical learning method based on reward and punishment \cite{Marco:12}. The agent interacts with an environment through its observations, actions and a scalar reward signal received from the environment. The agent's aim is to select actions that maximizes the cumulative future reward. Given a state, when an action increases (decreases) the value of an objective function (reward), which defines the goals of the agent, the probability of taking that action increases (decreases). Reinforcement learning algorithms involve estimating the state-action value function, or ``the Q value", which is a measure of how valuable, in terms of maximizing the total accumulated rewards, taking an action is, given the agent's state. This estimation is generally conducted in an iterative fashion by updating these Q values after each training step. In classical Q-learning reinforcement learning algorithm, for discrete state and action spaces, the update rule is given as \cite{Rie:05}, \cite{Rie:11}: 
\begin{equation}
Q(s,a)\rightarrow(1-\alpha)Q(s,a)+\alpha(r(s,a,s')+\gamma\max\limits_{a}Q(s',a))
\end{equation}
where, $s$, $a$, and $s'$ refer to the state, action, and the successor state, respectively. $\alpha$ is a learning rate and $\gamma$ is a discounting factor. The discounting factor emphasizes the relative importance of future rewards compared to current rewards. Neural Fitted Q Iteration (NFQ) method \cite{Rie:05}, \cite{Rie:11} approximates the Q value function using a neural network of type multi-layer perceptron. For a given state-action pair the neural network takes the state and action as its input and provides an approximate value of the corresponding Q value for the state-action pair. The method's aim is to minimize the following error function \cite{Rie:05}

\begin{equation}
(Q(s,a)-(r(s,a,s')+\gamma\max\limits_{a}Q(s',a)))^{2}.
\end{equation}
where, $r(.)$ is the reward signal that agent receives from the environment after the transition from state $s$ to state $s'$ by taking action $a$. This error function measures the deviations between state-action Q values approximated by the multi-layer perceptron ($Q(s,a)$) and the target value ($r(s,a,s')+\gamma\max\limits_{a}Q(s',a))$). In the NFQ method, Q value functions are updated in batches meaning that the entire set of input patterns ($(s_{i},a_{i}), i=0,1,2,3...$) and target patterns ($r(s_{i},a_{i},s_{i}')+\gamma\max\limits_{a}Q(s_{i}',a_{i})), i=0,1,2,3...$) are collected and the update is performed at the end of a full episode. To summarize, the NFQ method consists of two major steps: the generation of a training set and training a multi-layer perceptron using this set to obtain a Q-value function approximating the optimal state-action Q-values, at the end of each episode. The training is stopped whenever the received average reward per episode converges.

The goal of the reinforcement learning algorithm is to learn the optimal Q values by maximizing the agent's return, which is calculated via a reward/objective function. A reward function can be considered as a happiness function, goal function or utility function which represents, mathematically, the preferences of the pilot. In this paper, the pilot reward function is defined as
\begin{multline} 
reward = w1*(-C)+w2*(-S)+w3*(-A)+w4*(-P).
\end{multline}
In (11) $``C"$ is the number of aircraft within the collision region. Based on the definition provided by the Federal Aviation Administration (FAA), the radius of collision is taken as $500ft$ in the horizontal direction and $100ft$ in the vertical direction \cite{NextGen:07}. $``S"$ is the number of air vehicles within the separation region. The radius of the separation region is $5nm$ in the horizontal direction \cite{Perez:12} and $1000ft$ in the vertical direction based on the ``Reduced vertical separation minima" \cite{NextGen:07}. $``A"$ represents whether the aircraft is getting closer to the intruder or going away from the intruder in terms of their approach angle and takes the values of 1, for getting closer, or 0, for going away. $``P"$ represents whether the aircraft gets closer to or goes away from its trajectory vector in terms of angle and takes the values of 0, for getting closer, or 1, for going away. 

\begin{figure}[!b]
	\centering	
	\includegraphics[width=8cm]{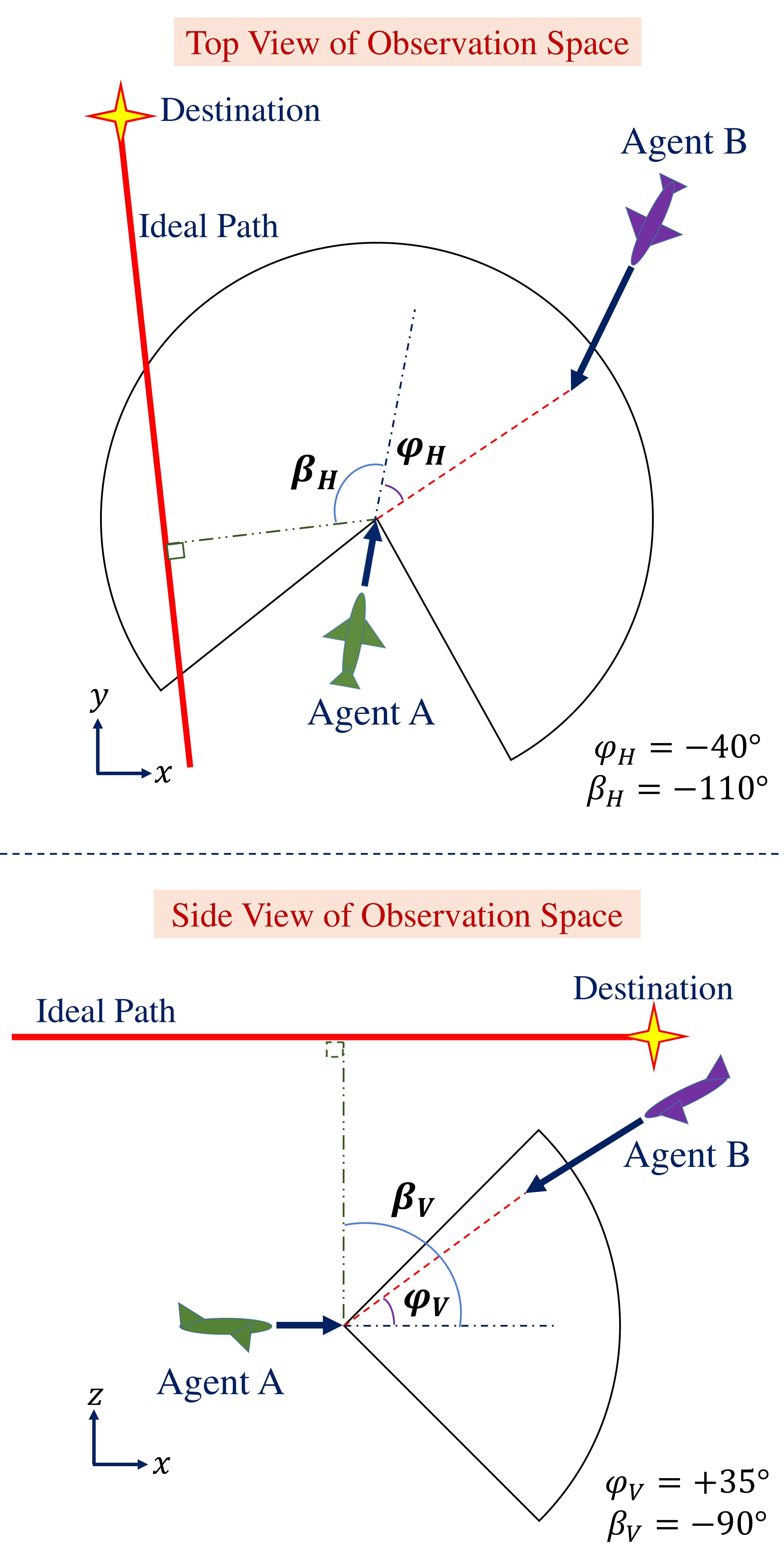}
	\caption{Pilot observation space.}
	\label{f:ObservationSpace}
\end{figure}

\begin{figure}[htb]
	\centering	
	\includegraphics[width=8cm]{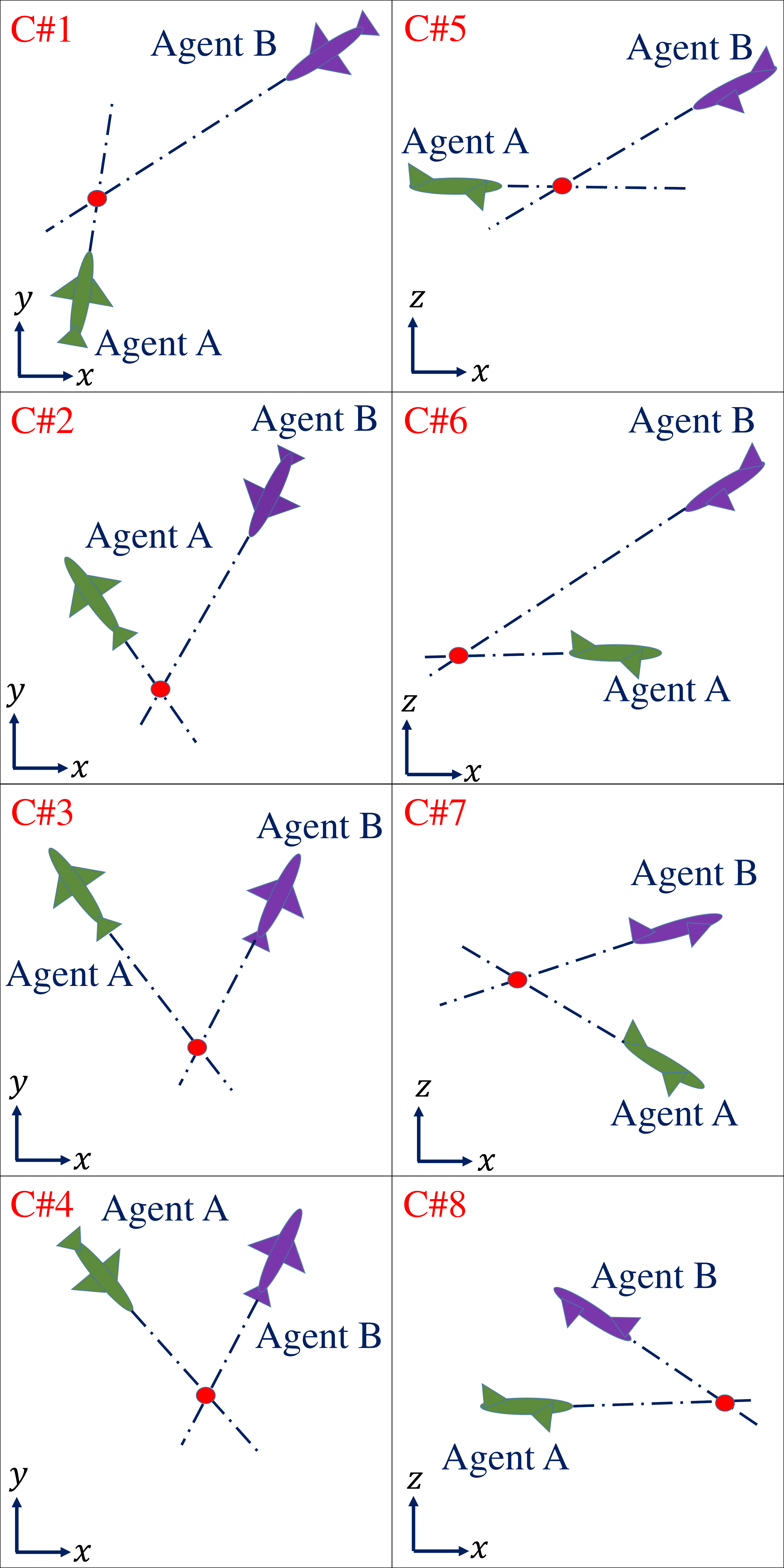}
	\caption{Encounter types.}
	\label{f:EC}
\end{figure}

Although ADS-B provides the positions and the velocities of other aircraft, with his/her limited cognitive capabilities a pilot can not possibly process all this information during his/her decision making process. In this study, in order to model pilot limitations, including the limitations at visual acuity and perception depth, as well as the limited viewing range of an aircraft, it is assumed that the pilots can observe (or process) the information from a limited portion of the nearby airspace. This limited portion is called the ``observation space". Since the aircraft are moving in a 3D region, the observation space is a 3D portion of the nearby airspace. This observation space is considered as a portion of a sphere centered at the location of the pilot. In order to illustrate the observation space, it is divided into horizontal and vertical parts and is schematically depicted in Fig.~\ref{f:ObservationSpace}. Viewing range of a pilot may be different in horizontal and vertical directions, which is why the observation space in these two directions are shown as different angular portions of a circle. Since the standard separation for manned aviation is $3-5nm$ \cite{Perez:12}, the radius of observation space is taken as a variable larger than $5nm$. Whenever an intruder aircraft moves toward the observation space (see Fig.~\ref{f:ObservationSpace}, where Agent B is the intruder), the approach geometry is defined by two angles: $\phi_{H}$, in the horizontal plane, and $\phi_{V}$, in the vertical plane. Aircraft's angular orientation with respect to his/her ideal trajectory is also defined by two angles: $\beta_{H}$, in the horizontal plane, and $\beta_{V}$, in the vertical plane. Fig.~\ref{f:ObservationSpace} depicts a typical example, where the aircraft B is moving toward the observation space with $\phi_{H}=-40^{\circ}$, $\phi_{V}=+35^{\circ}$, $\beta_{H}=-110^{\circ}$ and $\beta_{V}-90^{\circ}$. Aircraft relative orientations are also coded as different ``encounter types". Fig.~\ref{f:EC} depicts 8 types of encounter geometries projected in the horizontal plane (left column) and the vertical plane (right column). These geometries are indicated as $C\#i, i=1, 2, ..., 8$ in the figure. Finally, the observation space includes the pilot's memory of what their actions were at the previous time step. Given an observation, the pilots can choose between five actions: turn $45^{\circ}$ left, go straight, turn $45^{\circ}$ right, pitch $10^{\circ}$ up, or pitch $10^{\circ}$ down. It is noted that these pilot commands are filtered through the aircraft dynamics provided in Section~\ref{Manned Aircraft}.  


The information of the observations and actions of the pilots is fed to the neural network which is in charge of approximating the Q values. The input vector fed into the neural network is $[sign(\beta_{H})$, $sign(\beta_{V})$, $intruder status$, $sign(\phi_{H})$, $sign(\phi_{V})$, $encounter type$, $previous action$, $action]^{T}$, where $sign(.)$ takes the values of $+1$, $-1$ or $0$ depending on whether its argument value is positive, negative or zero, respectively. The $intruder status$ is taken as $1$ whenever an intruder is detected by the pilot, and $0$, otherwise. The $encounter type$ (see Fig.~\ref{f:EC}) is fed to the neural network in the form of a vector with 2 elements indicating the encounter type in horizontal plane and vertical plane. The 1st element takes the values of $-1$, $-0.5$, $0.5$ and $+1$ for encounter types of $C\#1$, $C\#2$, $C\#3$ and $C\#4$ in horizontal plane, and $0$, otherwise. Similarly, the 2nd element takes the values of $-1$, $-0.5$, $0.5$ and $+1$ for encounter types of $C\#5$, $C\#6$, $C\#7$ and $C\#8$ in vertical plane, and $0$, otherwise. The $previous action$ and $action$ are fed to the neural network in the form of a vector with 4 elements: actions ``turn $45^{\circ}$ left", ``go straight", ``turn $45^{\circ}$ right", ``pitch $10^{\circ}$ up", or ``pitch $10^{\circ}$ down" are coded as $[0,0,0,1]$, $[0,0,1,0]$, $[0,1,0,0]$, $[1,0,0,0]$ and $[-1,-1,-1,-1]$, respectively. It is noted that to improve the precision, instead of using the signs of the orientation angles their continuous values can be used.

\section{Simulation Results and Discussion}\label{Simulation Results and Discussion}
In this section, a quantitative analysis of multiple UAS integration in a crowded airspace is presented. Before presenting these results, single encounter scenarios, where two manned aircraft with different reasoning levels which are in a collision path, are investigated.

\subsection{Single Encounter Scenarios for manned aircraft}\label{Scenarios between manned aircraft}
\begin{figure}[tb]
	\centering	
	\includegraphics[width=10cm]{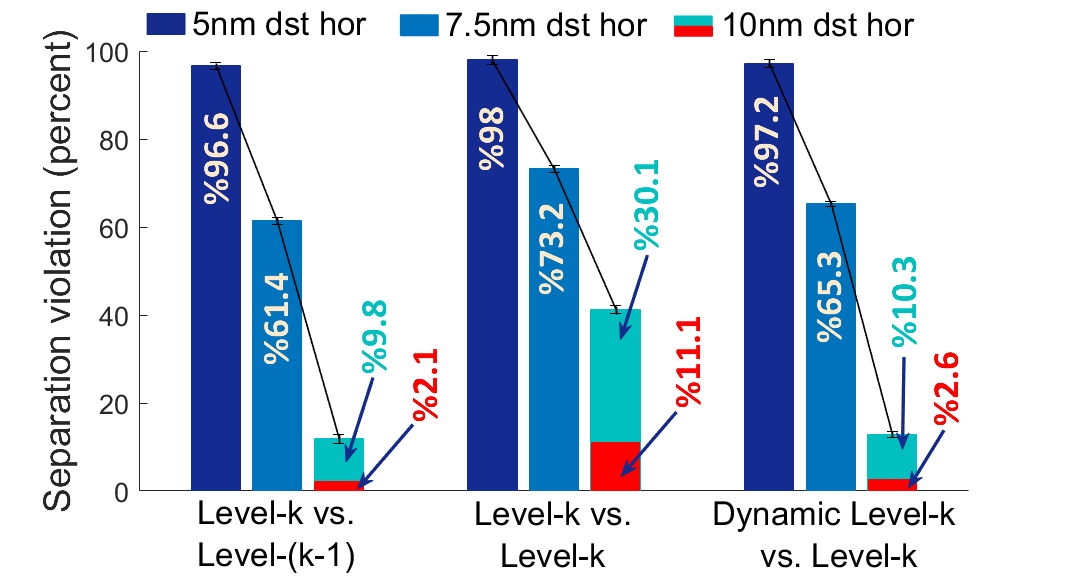}
	\caption{Separation violation rates. ``dst hor" refers to the distance horizon of the pilots. On the columns that show the separation violation rates for a $10nm$ distance horizon, the cyan color shows the percentage of violations that occur when the encounters are ``difficult to resolve", meaning that the initial conditions of the encounters do not permit any type of pilot action to avoid a separation violation.}
	\label{f:pblevel}
\end{figure}

Fig.~\ref{f:pblevel} presents the separation violation rates of manned aircraft in 5000 random single encounters where pilots are modeled as level-1 and level-2 decision makers. Separation violation occurs when the horizontal and the vertical distances between the two aircraft are less than the horizontal separation requirement, $5nm$ \cite{Perez:12} and the reduced vertical separation requirement, $1000ft$ \cite{Perez:12}, respectively. In the figure, separation violations are shown for 3 different ``distance horizon" values, which is the radius of the observation space depicted in Fig.~\ref{f:ObservationSpace}. Pilots oversee a $20s$ time window prior to a probable separation violation with a $5second$ decision frequency for choosing their actions. Distance horizon takes three values: $5nm$ (equal to horizontal separation requirement), $7.5nm$ and $10nm$. Pilots are either level-0, level-1 or level-2 agents. There are 4 possible types of scenarios: 1) level-1 pilot vs. level-0 pilot, 2) level-2 pilot vs. level-1 pilot, 3) level-1 pilot vs. level-1 pilot, and 4) level-2 pilot vs. level-2 pilot. According to Fig.~\ref{f:pblevel}, in the 1st type and 2nd type scenarios (level-k vs level-(k-1)), by increasing the distance horizon from $5nm$ to $10nm$ the separation violation rate decreases from $96.6\%$ to $11.9\%$. It is noted that $82\%$ (9.8/11.9 = 0.82) of the $10nm$ distance horizon separation violations occur during the scenarios where the encounters are difficult to be resolved. This is because for these type of encounters, no matter how pilots in the collision path maneuver to resolve the conflict the separation violation occurs. In the 3rd and 4th type of scenarios (level-k vs level-k) the separation violation rate decreases by increasing the pilots' distance horizon, however, separation violation rate is still high ($41.2\%$) even for the $10nm$ case. The reason for this high separation violation rate is that when a level-k agent encounters another agent with the same level, his/her assumption about the other becomes invalid. This problem and its implications were discussed in Section \ref{Dynamic Level-k Reasoning}, where a remedy was proposed (see Fig~\ref{f:PSC}), which was termed as ``dynamic" level-k reasoning method. According to Fig.~\ref{f:pblevel}, the separation violation rate, for regular encounters (red), decreases from $11.1\%$ to $2.6\%$, when dynamic level-k reasoning is chosen as the interactive decision making model, for the case of $10nm$ distance horizon.

To visually demonstrate the pilot decision making process, two 3D single encounter scenarios are designed. The first scenario consists of two manned aircraft flying towards each other at different altitude levels, where the initial horizontal and vertical distances between the aircraft are $21.3nm$ and $1000ft$, respectively. The 2nd scenario consists of two manned aircraft flying towards each other where one of them is flying on a horizontal plane while the other is leveling up with a constant vertical rate of $1968ft/min$. In the second scenario, initial horizontal and vertical distances between the aircraft are $21nm$ and $1000ft$, respectively. In the scenarios, pilots can oversee conflicts in a $20s$ time window prior to a separation violation. Pilots' distance horizon is considered to be $10nm$. Both of the aircraft can change the heading angle for $\pm45^{\circ}$, or change the pitch angle for $\pm10^{\circ}$, or may keep both the heading and pitch angles unchanged. In both of the scenarios, 4 cases for pilots' level-k type behavior are considered: 1) a level-1 pilot vs. a level-0 pilot, 2) a level-2 pilot vs. a level-1 pilot, 3) a level-1 pilot vs. a level-1 pilot, and 4) a dynamic level-1 pilot vs. a level-1 pilot. 

\begin{figure}[tb]
	\centering	
	\includegraphics[width=8cm]{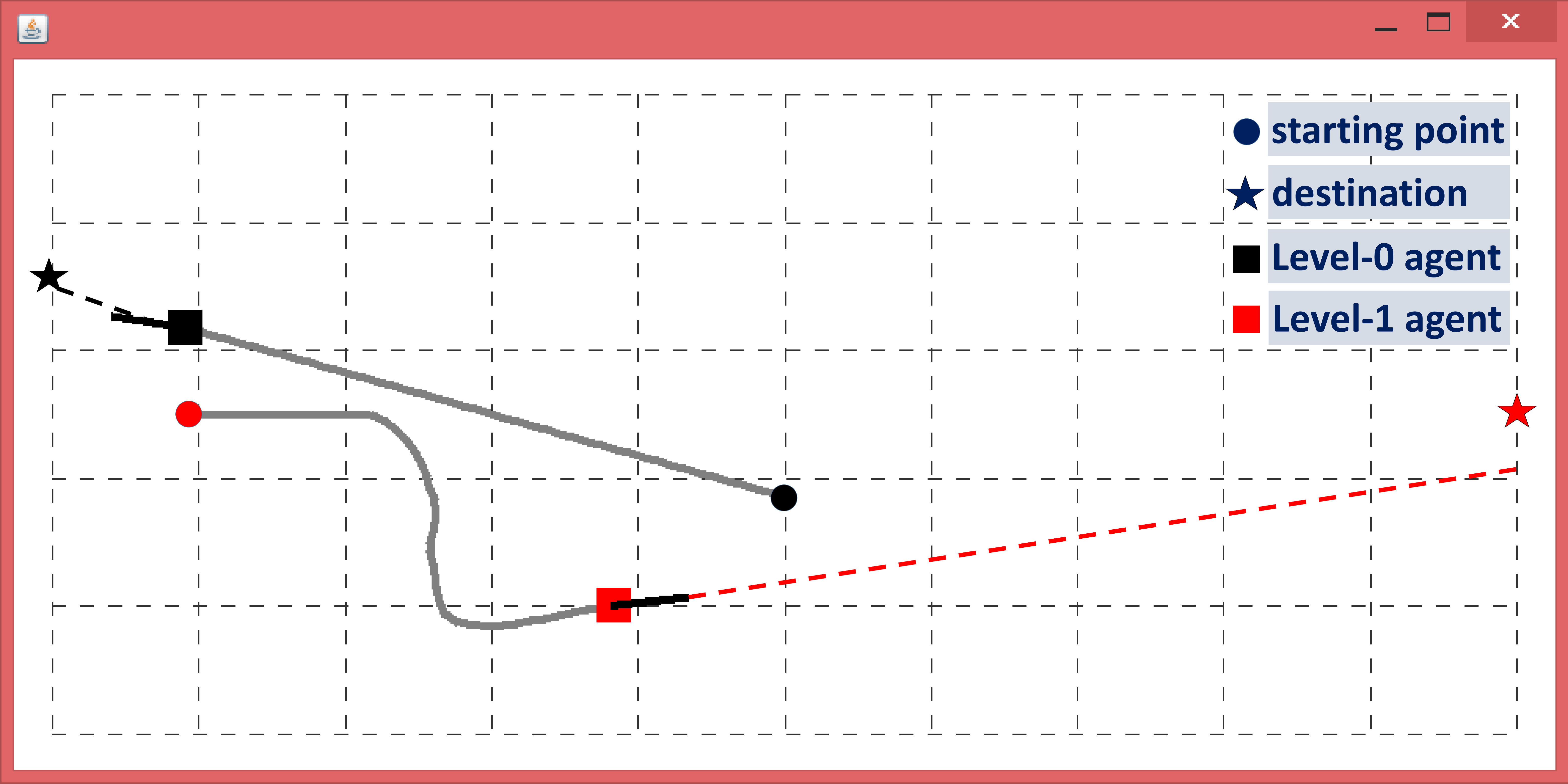}
	\caption{Encounter scenario 1: level-1 pilot vs. level-0 pilot.}
	\label{f:S1}
\end{figure}

\begin{figure}[htb]
	\centering	
	\includegraphics[width=8cm]{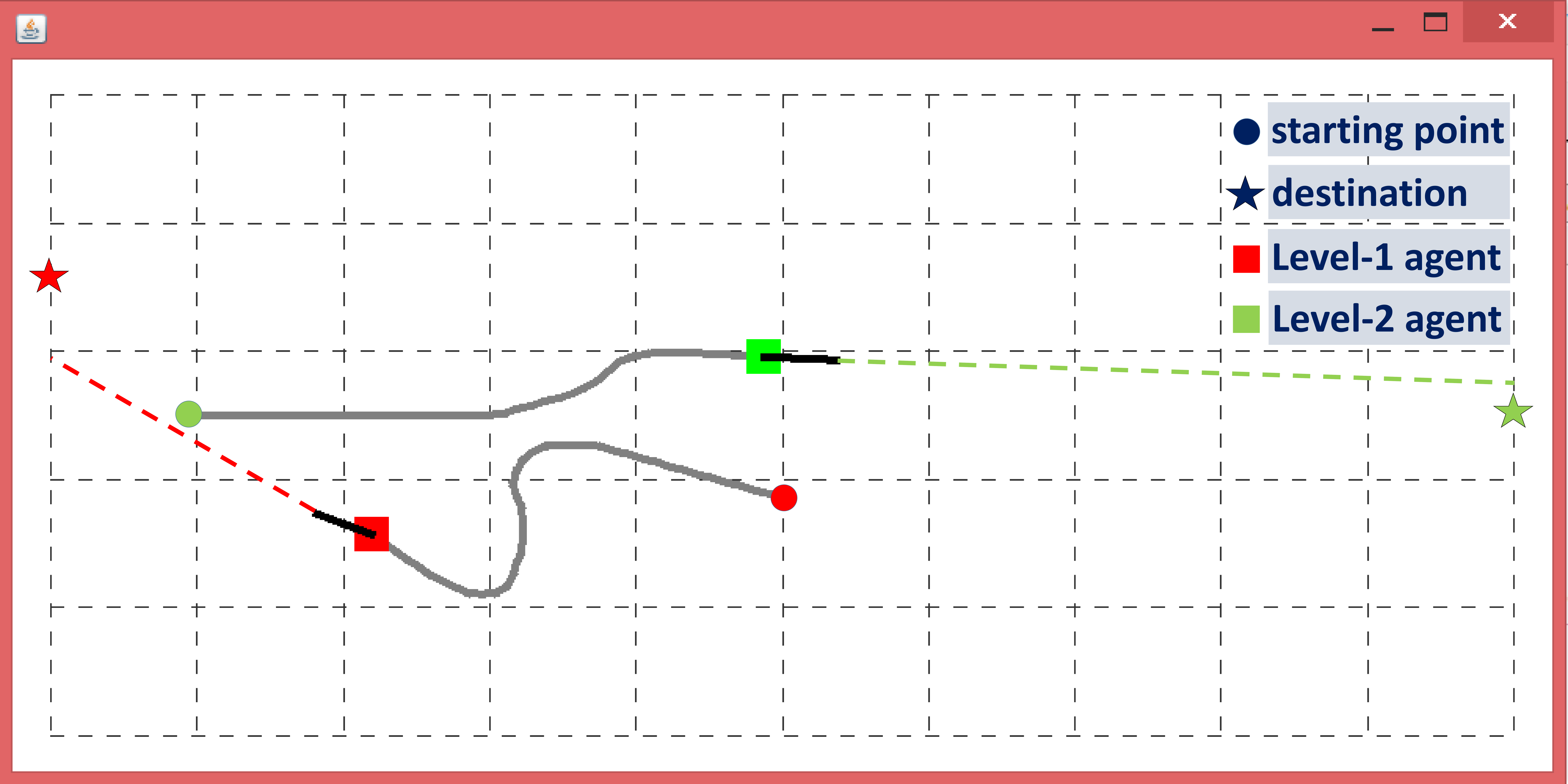}
	\caption{Encounter scenario 1: level-2 pilot vs. level-1 pilot.}
	\label{f:S2}
\end{figure}

\begin{figure}[htb]
	\centering	
	\includegraphics[width=8cm]{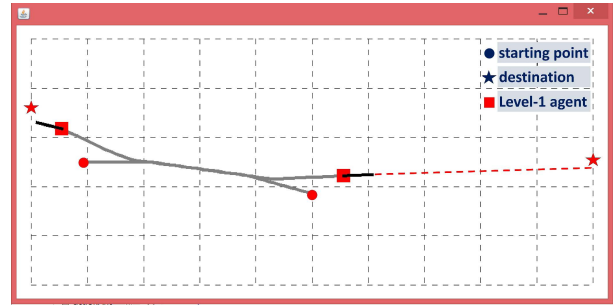}
	\caption{Encounter scenario 1: level-1 pilot vs. level-1 pilot.}
	\label{f:S3}
\end{figure}

\begin{figure}[!htb]
	\centering	
	\includegraphics[width=8cm]{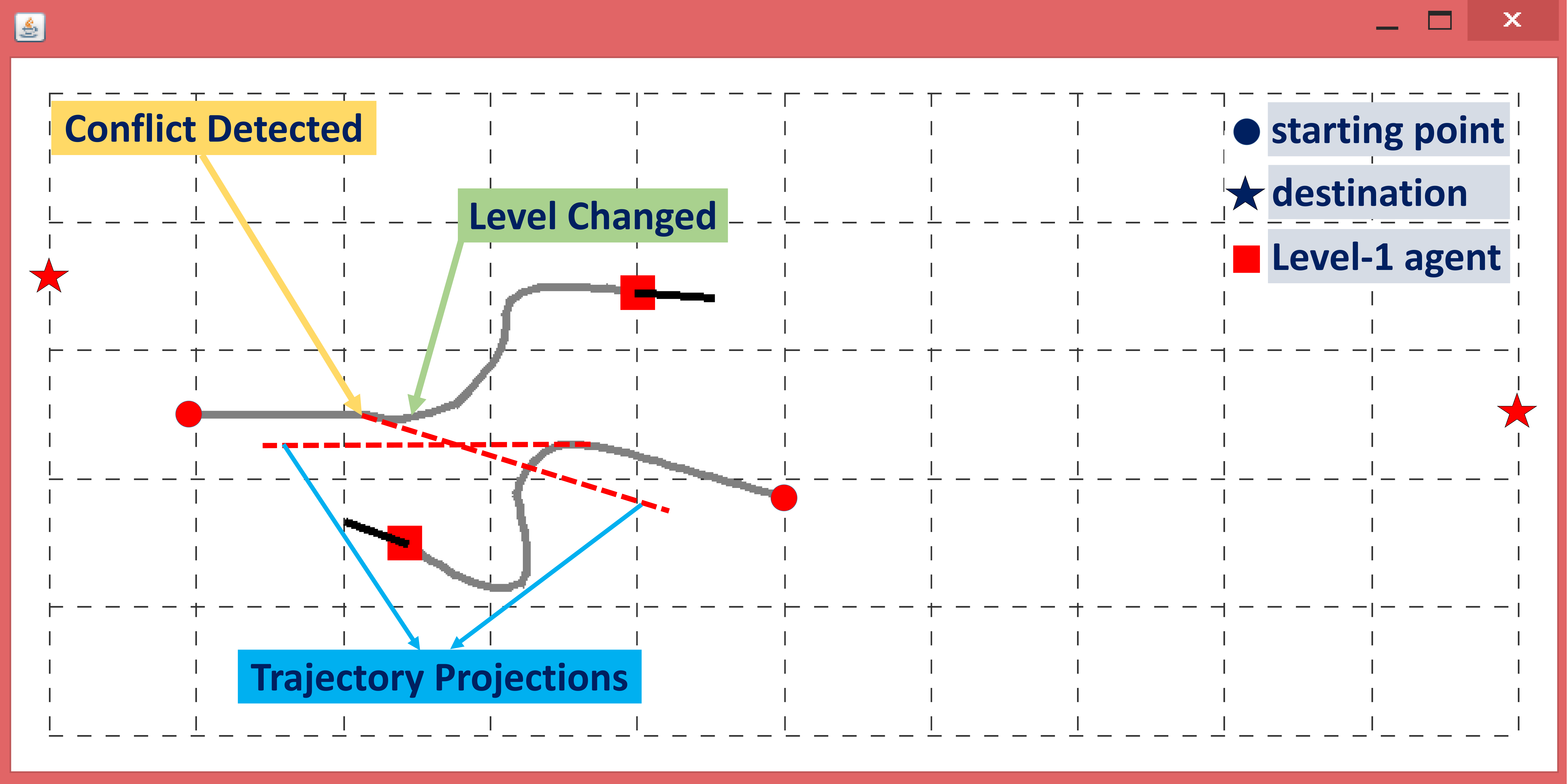}
	\caption{Encounter scenario 1: dynamic level-k pilot vs. level-1 pilot.}
	\label{f:S4}
\end{figure}

\begin{figure}[htb]
	\centering	
	\includegraphics[width=8cm]{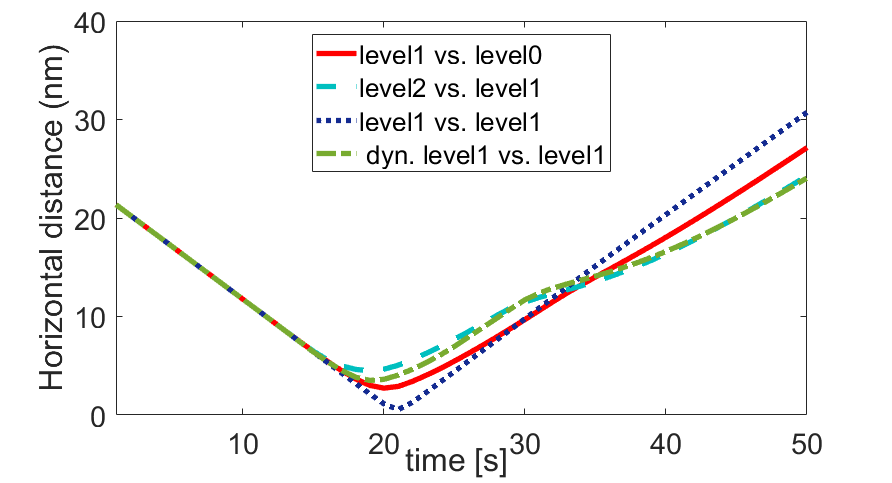}
	\caption{Encounter scenario 1: horizontal distance.}
	\label{f:S5}
\end{figure}

\begin{figure}[htb]
	\centering	
	\includegraphics[width=8cm]{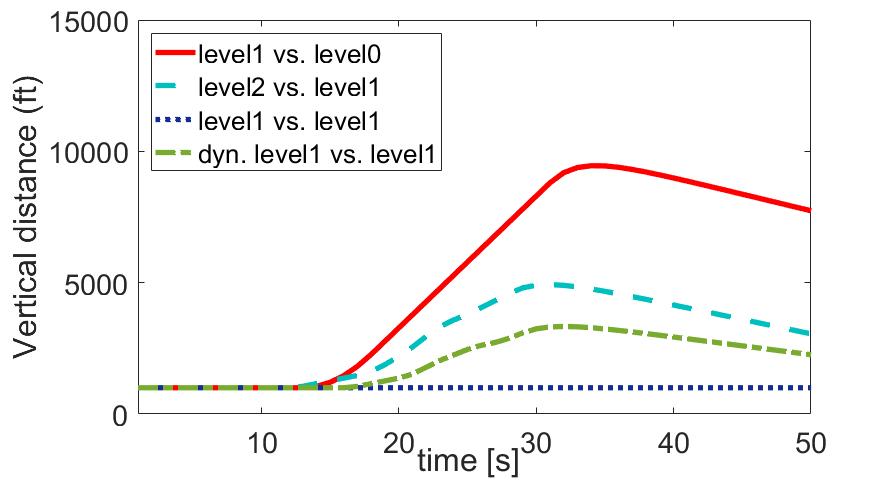}
	\caption{Encounter scenario 1: vertical distance.}
	\label{f:S6}
\end{figure}

Figures~\ref{f:S1}-\ref{f:S4} depict the 2D horizontal projection snapshots of the four cases, for the first scenario. Black, red and green squares correspond to manned aircraft with level-0, level-1 and level-2 type pilots. The circles and stars stand for the initial positions and final destinations of the aircraft. The gray track lines right behind the aircraft represent their traveled path from their initial positions to where they stand in the snapshot. Two neighboring
grid points are $5nm$ away. It is seen from these figures that while level-1 vs level-0 and level-2 vs level-1 encounter conflicts are properly resolved, level-1 vs level-1 conflict resulted in a separation violation. As explained earlier, this problem might occur in certain approach geometries due to incorrect opponent level assumption. Figure 9 shows an encounter scenario where a dynamic level-k pilot has an encounter with a level-1 pilot. The dynamic level-k pilot starts with a level-1 policy and when he/she detects a probable conflict, changes his/her policy to level-2 and prevents a separation violation. Figure~\ref{f:S5} and Fig.~\ref{f:S6} depict the horizontal and vertical distances between the two aircraft during the simulation of four cases discussed here. It is seen that no separation violation occurs except the level-1 vs level-1 case, for the 1st scenario.

\begin{figure}[b]
	\centering	
	\includegraphics[width=8cm]{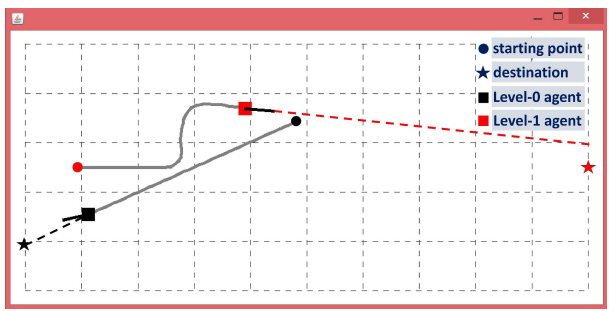}
	\caption{encounter scenario 2: level-1 pilot vs. level-0 pilot.}
	\label{f:S7}
\end{figure}

\begin{figure}[htb]
	\centering	
	\includegraphics[width=8cm]{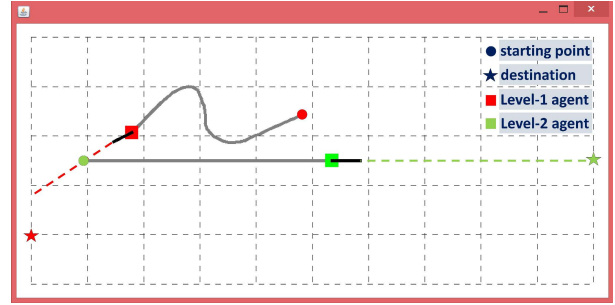}
	\caption{Encounter scenario 2: level-2 pilot vs. level-1 pilot.}
	\label{f:S8}
\end{figure}

\begin{figure}[htb]
	\centering	
	\includegraphics[width=8cm]{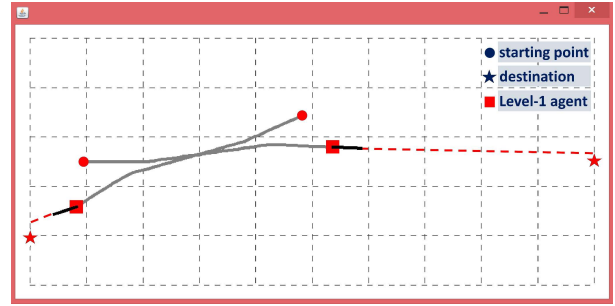}
	\caption{Encounter scenario 2: level-1 pilot vs. level-1 pilot.}
	\label{f:S9}
\end{figure}

\begin{figure}[!htb]
	\centering	
	\includegraphics[width=8cm]{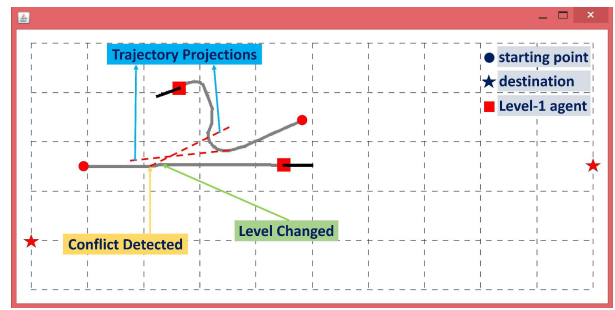}
	\caption{Encounter scenario 2: dynamic level-k pilot vs. level-1 pilot.}
	\label{f:S10}
\end{figure}

Similar observations can be drawn for the simulations of the second scenario, the results of which are presented in Figures~\ref{f:S7}-\ref{f:S10}.

\begin{figure}[htb]
	\centering	
	\includegraphics[width=8cm]{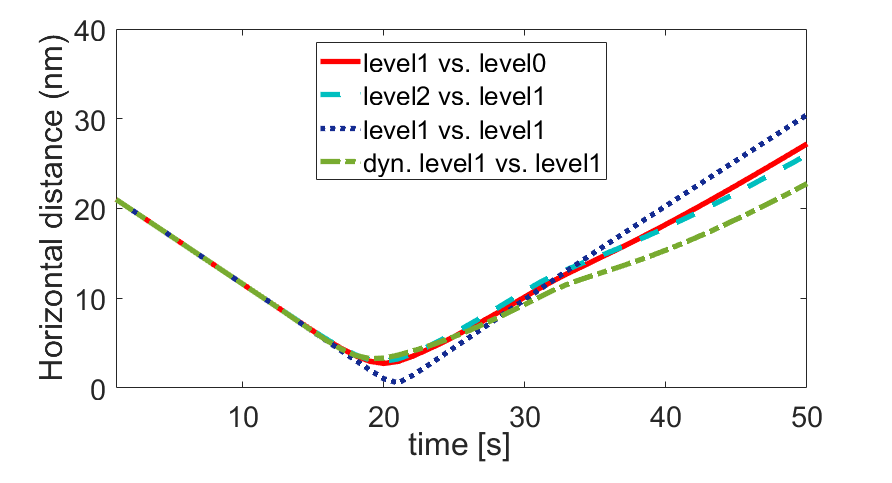}
	\caption{Encounter scenario 2: horizontal distance.}
	\label{f:S11}
\end{figure}

\begin{figure}[htb]
	\centering	
	\includegraphics[width=8cm]{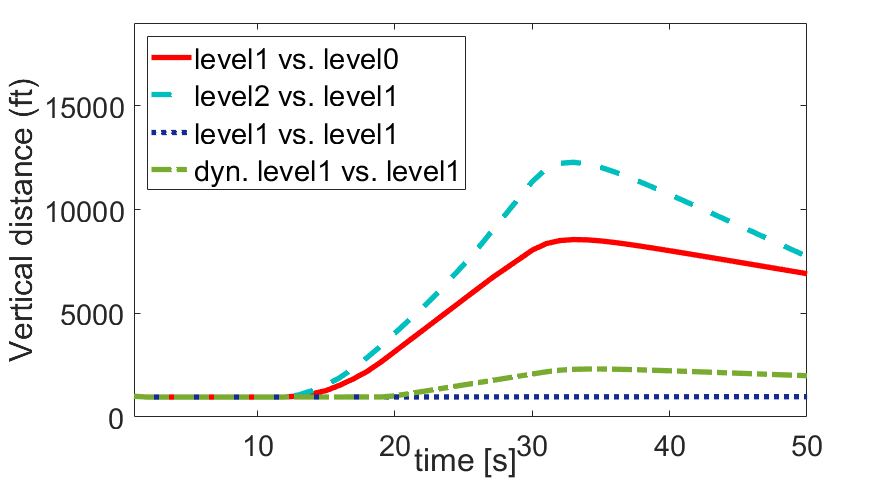}
	\caption{Encounter scenario 2: vertical distance.}
	\label{f:S12}
\end{figure}

\subsection{UAS flying in a crowded airspace}\label{Statistical Results For UAS Integration}
In this section, the scenario explained in section \ref{UAS Integration Scenario} is simulated to investigate the effect of \textit{responsibility assignment} for conflict resolution, on safety and performance. Separation responsibility assignment is an important issue in addressing the integration of UAS into NAS \cite{NASA:12}: it is crucial to determine which of the agents (manned aircraft or UAS) will take the responsibility of the conflict resolution. Since the loss of separation is the most serious issue, the safety metric is taken as the total number of separation violations between all aircraft whether manned or unmanned. Performance metric, on the other hand, is taken as the averaged manned and unmanned aircraft trajectory deviations. In all of the simulations, level-0, level-1 and level-2 pilot policies are randomly distributed over the manned aircraft in such a way that 10\% of the pilots fly based on level-0 policies, 60\% based on level-1 policies and 30\% based on level-2 policies. This distribution is obtained from human experimental studies discussed in \cite{Costa:95} and may not necessarily reflect the true distribution for the scenarios discussed here. It is noted, however, that level distribution can be adapted to other distributional data in the proposed framework. Level-1 type and level-2 type pilots utilize the dynamic level-k reasoning method.

\begin{figure}[b]
	\centering	
	\includegraphics[width=8cm]{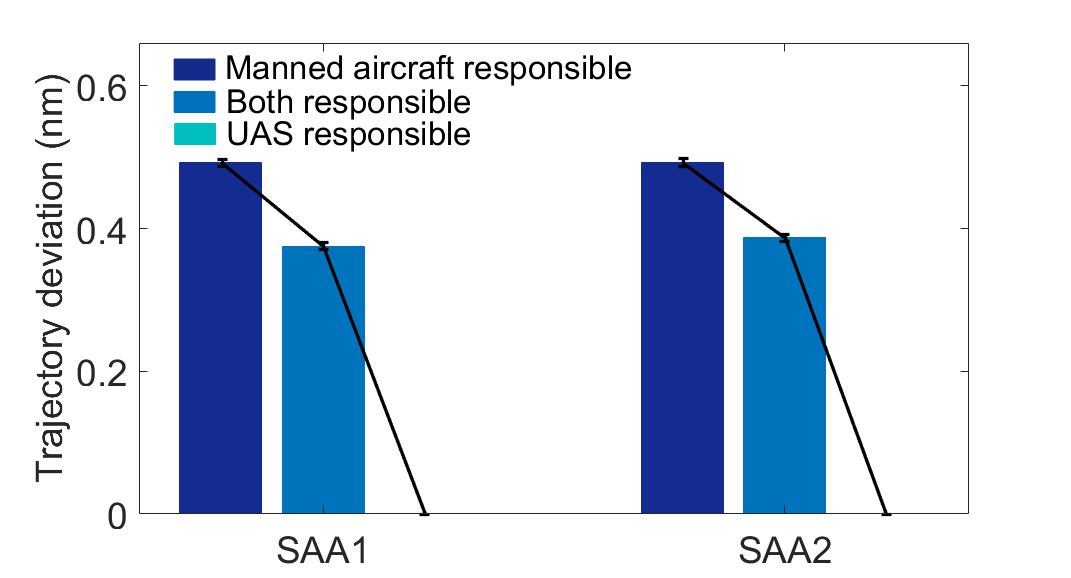}
	\caption{Average trajectory deviation of manned aircraft.}
	\label{f:R1}
\end{figure}

\begin{figure}[tb]
	\centering	
	\includegraphics[width=8cm]{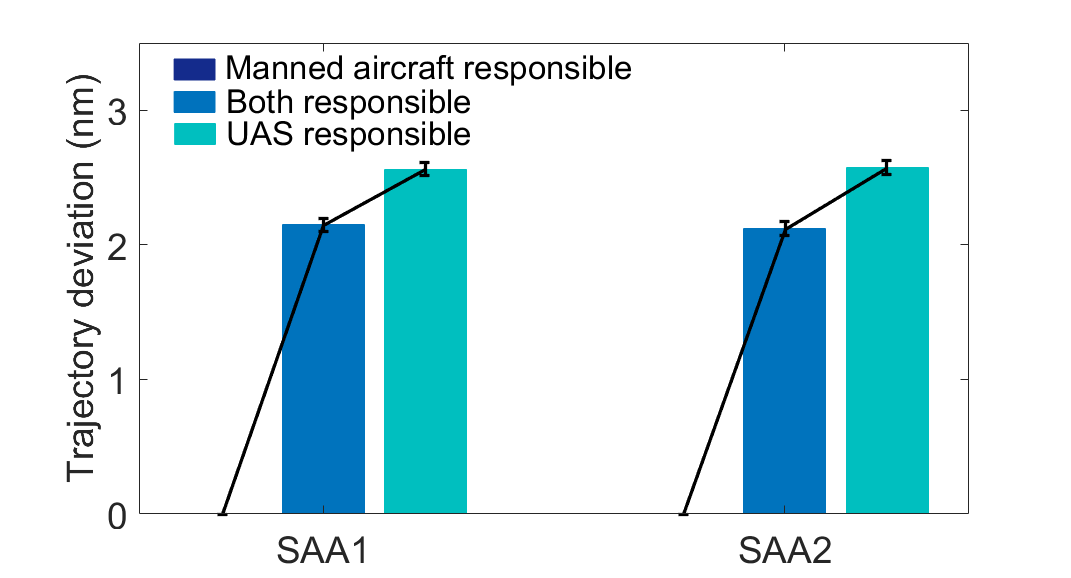}
	\caption{Average trajectory deviation of UAS.}
	\label{f:R2}
\end{figure}

\begin{figure}[tb]
	\centering	
	\includegraphics[width=8cm]{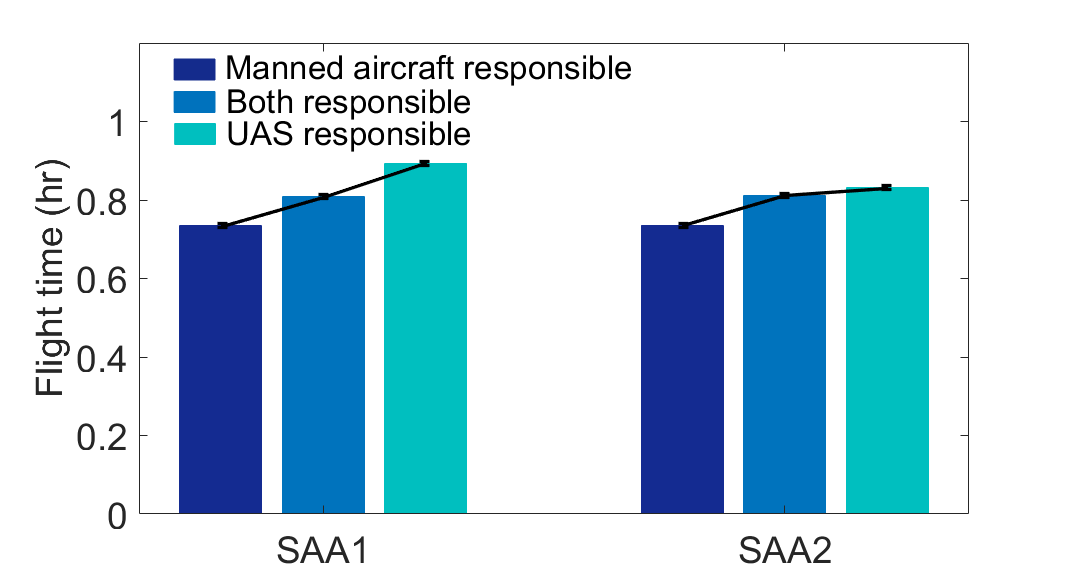}
	\caption{UAS flight time.}
	\label{f:R3}
\end{figure}

\begin{figure}[!ht]
	\centering	
	\includegraphics[width=8cm]{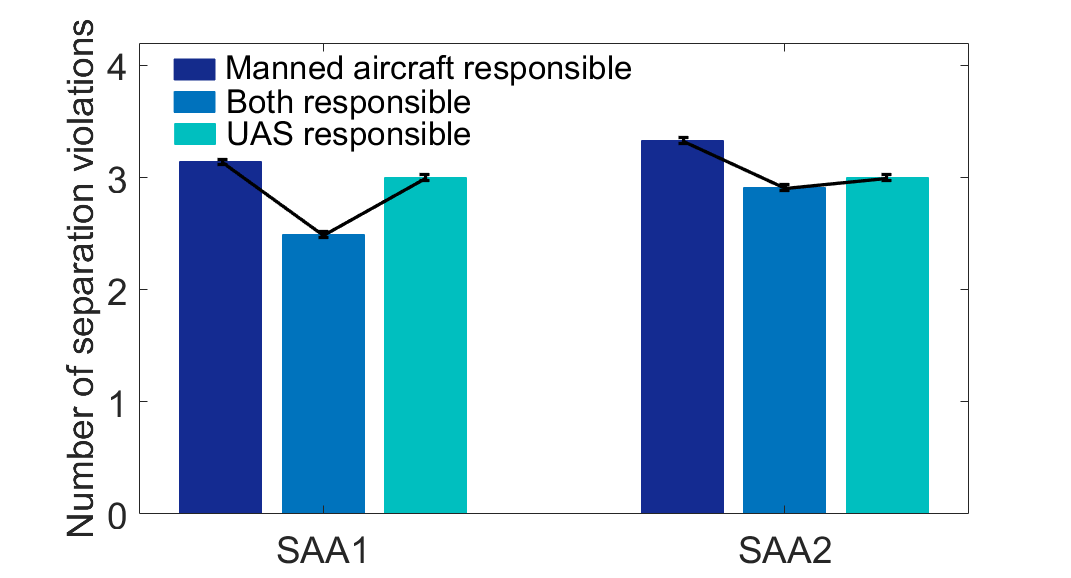}
	\caption{Separation violation between manned aircraft and UAS.}
	\label{f:R4}
\end{figure}

Fig.~\ref{f:R1}, Fig.~\ref{f:R2}, Fig.~\ref{f:R3}, and Fig.~\ref{f:R4} depict a comparison of different resolution responsibility cases: manned aircraft are responsible (dark blue), both manned aircraft and UAS are responsible (blue) and UAS are responsible (cyan). Both SAA1 and SAA2 logics are employed in the simulations. In the case when only manned aircraft are responsible for conflict resolution, UAS are forced to continue their path without running their SAA system and manned aircraft act as dynamic level-1 and level-2 decision makers. In the case when the UAS are responsible for the conflict resolution, the manned aircraft are forced to continue their path without changing their heading and the UAS execute the maneuvers dictated by their SAA algorithms. In the case when both the manned aircraft and the UAS are responsible for the conflict resolution, they both utilize their evasive maneuvers. Fig.~\ref{f:R1} shows that manned aircraft deviate more from their trajectories when only the manned aircraft share resolution responsibility, compared to the case when both the UAS and the manned aircraft are responsible. This is true for both the SAA1 (on the left) and the SAA2 (on the right) algorithms. 
On the other hand, Fig.~\ref{f:R2} shows that the UAS deviates from its trajectory more when it is responsible for the resolution, compared to the case when the responsibility is shared, which holds true for both of the SAA methods. Figure~\ref{f:R3} shows, as expected, that for both SAA1 and SAA2, the UAS flight times are the shortest when only the manned aircraft become responsible for the resolution. According to Fig.~\ref{f:R4}, for both SAA1 and SAA2 utilizations, the minimum number of separation violations are observed when conflict resolution responsibility is shared between the UAS and the manned aircraft. 

\section{Conclusion}\label{Conclusion}
In this paper, a combination of the level-k reasoning game theoretical concept and an approximate reinforcement learning method called Neural Fitted Q-learning is used to create a 3-dimensional (3D) airspace modeling framework for predicting the possible outcomes of integrating Unmanned Aircraft Systems (UAS) into the National Airspace System (NAS). Compared to the earlier results of the authors, the assumption that the decision makers' levels remain the same during interactions and the requirement of keeping a large Q-value table are removed. These are achieved by the introduction of a dynamic level-k reasoning method and the employment of the Neural Fitted Q-learning algorithms, respectively. These improvements made it possible to model a larger class of interactions between the decision makers and this is demonstrated by simulating various single encounter scenarios in a 3D airspace. The proposed modeling framework can be used to quantitatively investigate how safety and performance of the simulated airspace system are affected by the various integration technologies and concepts such as airspace density, minimum separation distance and various UAS sense and avoid algorithms and their design parameters. One of the issues about UAS integration is responsibility assignment during conflicts and it is shown how the 3D game theoretical modeling framework discussed in this paper can be used to study this problem.

\section*{Acknowledgment}
This effort was sponsored by the Scientific and Technological Research Council of Turkey under grant number 114E282.


\end{document}